\documentstyle[seceq,epsf,preprint]{ptptex}

\preprintnumber{
OU-TAP 48
}
\title{
The Spectrum of Gravitational Wave Perturbations 
in the One-Bubble Open Inflationary Universe
}

\author{
Takahiro {\sc Tanaka}\footnote{Electronic address: 
tama@vega.ess.sci.osaka-u.ac.jp} 
and Misao {\sc Sasaki}\footnote{Electronic address: 
misao@vega.ess.sci.osaka-u.ac.jp}
}

\inst{
Department of Earth and Space Science, Osaka University, Toyonaka 560
}

\recdate{
\today
}

\abst{
We give the initial spectrum of quantized gravitational waves 
in the context of the one-bubble open inflationary universe scenario. 
In determining the quantum state after the bubble nucleation 
we adopt the prescription to require the analyticity 
of positive frequency functions in half of the 
Euclidean extension of the background 
$O(3,1)$-symmetric spacetime.
We find the spectrum is well behaved at the infrared limit
and there appears no supercurvature mode.
In the thin wall approximation, the explicit form of the 
spectrum of gravitational wave perturbations is calculated. 
}

\begin{document}

\maketitle

\section{Introduction}

The open inflationary universe scenario is one of exciting 
current topics among the physics in the early universe. 
The possibility to create an open universe $(\Omega_0<1)$ 
through bubble nucleation
in the context of inflationary cosmology has become under 
discussion rather recently,\cite{Got,BGT,YST95,Lindea,Lindeb,GreLid}
and several models of inflaton potential 
have been proposed.\cite{BGT,Lindea,Lindeb,GreLid}

Now a central issue is if these models are compatible with the 
observed anisotropies of cosmic microwave background 
(CMB) on large angular scales. In several recent
papers,\cite{YTS95,HAMA,STY95,ST96,YST96,YB,Garriga,Bellido,Cohn} 
quantum fluctuations of the inflaton field that generate the 
initial curvature perturbations have been evaluated and 
the resulting spectrum of CMB anisotropies has been calculated. 
But a drawback of all the previous studies is that
the gravitational degrees of freedom have not been 
taken into account. 

The incorporation of gravitational perturbations 
causes two effects. 
One is the coupling between perturbations of the inflaton field 
and those of the metric, which may alter the 
spectrum of the initial curvature perturbations drastically. 
The other is the contribution of
gravitational wave perturbations to the CMB anisotropy, 
which has not been taken into account at all in the previous analyses. 
As has been known, a constant time hypersurface in an 
open inflationary universe is not a Cauchy surface of the whole
spacetime.\cite{STY95}
Thus we cannot set commutation relations on this hypersurface 
when we consider quantization of a field in the open universe. 
This difficulty has been solved in the case of a scalar
field\cite{STY95,ST96} and 
recently a method to manage the gravitational wave modes 
in the Milne universe has been
 developed by the present authors\cite{TanSas} (Paper I). 

In this paper, we extend the result in Paper I
to make it applicable to a general open universe. 
The aim of this paper is to give the spectrum of 
gravitational wave perturbations in the context of
the open inflationary universe scenario. 

This paper is organized as follows. 
In section 2 we review the zeroth order approximation to an open 
universe model, which is based on the $O(4)-$symmetric bubble 
nucleation\cite{Col,ColDeL}, 
and explain the notation we use in the succeeding sections. 
In section 3 gravitational wave perturbations of 
the $O(4)-$symmetric bubble is investigated. 
First we show a similarity between the massless scalar field perturbation 
and the gravitational wave perturbation. 
Then using this similarity, we reinterpret the results for 
massless scalar field perturbations obtained in Ref.~\citen{ST96}
and give the spectrum of gravitational wave perturbations.
Section 4 is devoted to summary and discussion. 

In this paper, we use the units, $c=\hbar=1$, and adopt the metric
signature, $(-,+,+,+)$. 
\newpage

\section{Configuration of the $O(3,1)$-symmetric bubble}
We consider the Einstein scalar model with a single 
real scalar field which has a potential, $V(\sigma)$, 
as shown in Fig.~1. 
The Lagrangian is given by 
\begin{equation}
 {\cal L}=
  \sqrt{-g}\left[{1\over 2\kappa}R-{1\over 2}g^{\mu\nu}
  \partial_{\mu}\sigma \partial_{\nu}\sigma -V(\sigma)\right], 
\end{equation}
where $\kappa=8\pi G$, and
$g$ and $R$ are the determinant of the metric tensor 
and the curvature scalar, respectively. 

\begin{figure}
\centerline{\epsfbox{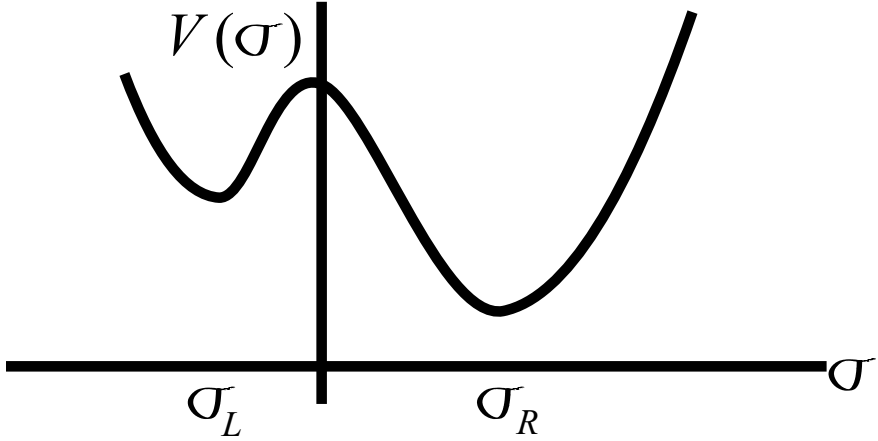}}
\caption{The potential of the scalar field}
\end{figure}

Here in this section, we review the background solution 
which represents the bubble nucleation.\cite{Col,ColDeL} 
The conformal diagram of this solution is presented in Fig.~2. 
As shown in Fig.~2, we divide the whole spacetime 
into five regions. 
The upper right and the upper left triangle regions 
are labeled by $R$ and $L$, respectively. 
The central, diamond-shaped region that
contains the bubble wall is labeled by $C$.
In $C$, the background metric is written as 
\begin{equation}
 ds_{C}^2 = 
  dT^2+a^2(T)
  \left(-dr_{C}^2+\cosh^2 r_{C} d\Omega^2\right),
\end{equation}
and the scalar field depends only on ``the cosmological time'', $T$,
\begin{equation}
 \sigma=\sigma(T). 
\end{equation}
Then the equations of motion become 
\begin{eqnarray}
 &&\ddot\sigma+3{\dot a\over a}\dot\sigma
   ={dV(\sigma)\over d\sigma},
\label{sigmaeq}
\\
 && \left({\dot a\over a}\right)^2
 -{1\over a^2}={\kappa\over 3}
 \left({1\over 2}\dot\sigma^2-V(\sigma)\right),  
\label{aeq}
\end{eqnarray}
where the dot $~\dot{~}~$ 
represents the derivative 
with respect to $T$. 

\begin{figure}[t]
\centerline{\epsfbox{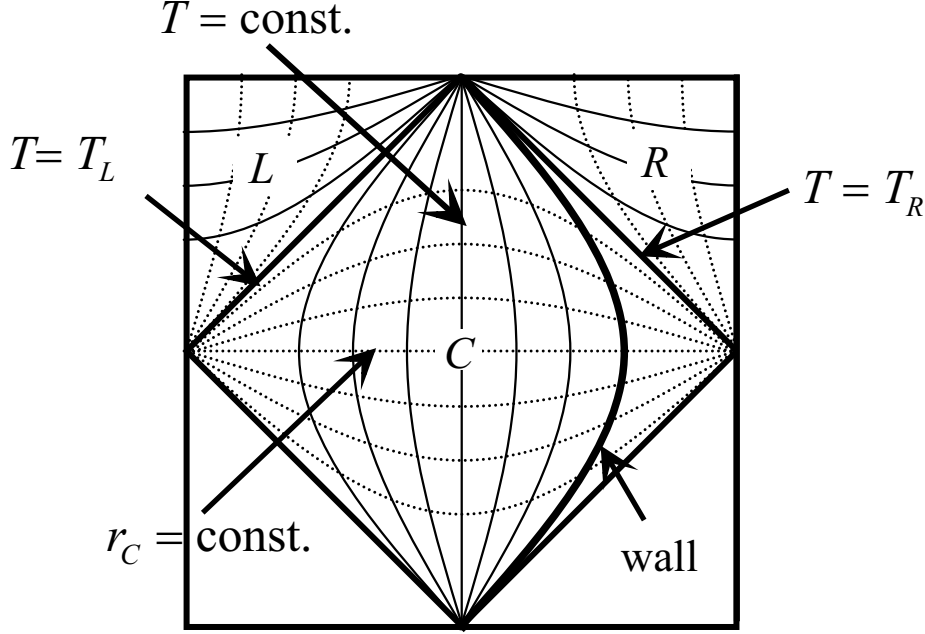}}
\caption{The conformal diagram of the universe 
containing an $O(3,1)$-symmetric bubble}
\end{figure}

The regularity of the metric requires 
the boundary condition for Eqs.~(\ref{sigmaeq}) and (\ref{aeq}) as
\begin{equation}
 \dot\sigma\vert_{T=T_J}=0,
\end{equation}
where $J=R$ or $L$ and 
$T_J$ is determined by $a(T_J)=0$. 
Hereafter the subscript $J$ is used to represent $R$ or $L$. 
The surface where $T=T_J$ corresponds to the boundary 
between $C$ and $J$.
As $T$ increases from $T_L$ to $T_R$, $\sigma$ also 
increases monotonically from $\sigma_L$ to $\sigma_R$. 
Here we assume that $\sigma_L$ is 
the potential minimum in the false vacuum side 
and that $\sigma_R$ 
is that in the true vacuum side.
Further, we assume that the region in which $\sigma$ 
changes are restricted to the interval between 
$T_{WL}$ and $T_{WR}$, where $T_{WL}<T_{WR}$. 
We divide the region $C$ into three regions;
$T<T_{WL}$, $T_{WL}<T<T_{WR}$ and $T_{WR}<T$, and 
call them $C_L$, $W$ and $C_R$, respectively. 
Thus $J$ and $C_J$ are the 
de Sitter space with $\sigma=\sigma_J$ with
the expansion rate given by $H_J:=\sqrt{\kappa V(\sigma_J)/3}$. 
As we shall see, these assumptions are not essential 
to our later discussion and they can be relaxed 
if one wishes. However, for simplicity, we consider the situation in
which these assumptions hold.

{}Following Ref.~\citen{ST96},
we introduce coordinates in $L$ and $R$. 
{}First we introduce 
new coordinates in $C_L$ and $C_R$ by 
\begin{equation}
   {dt_{C,J}\over H_J}=dT\raisebox{-5pt}{,}
    \quad a(T)={\cos t_{C,J} \over H_J}\raisebox{-5pt}{.}
\end{equation}
In these coordinates, the metric in $C_J$ takes the form,
\begin{equation}
 ds_{C,J}^2  = 
  H_J^{-2} \left[dt_{C,J}^2+\cos^2 t_{C,J}
  \left(-dr_{C}^2+\cosh^2 r_{C} d\Omega^2\right)\right].
\end{equation}
If we use the thin wall approximation, in which 
the region $W$ is assumed to be infinitesimally thin, the continuity 
of the metric gives the relation,
\begin{equation}
{\cos t_{C,L}\over H_L}\Bigr\vert_{wall}
={\cos t_{C,R}\over H_R}\Bigr\vert_{wall}\,\raisebox{-5pt}{.}
\end{equation}
Further we introduce the coordinates in $R$ and $L$ by 
\begin{eqnarray}
 t_R=i t_{C,R}-\pi i/2,\quad r_R=r_{C}+\pi i/2,
\cr
 t_L=-i t_{C,L}-\pi i/2,\quad r_L=r_{C}+\pi i/2.
\label{CoTr}
\end{eqnarray}
The relations among the coordinate systems are uniquely determined 
by the analyticity which was discussed in Ref.~\citen{YST96}. 
The metric in $R$ or $L$ becomes 
\begin{equation}
 ds_{J}^2 = 
  H_J^{-2}\left[ -dt_J^2+\sinh^2 t_J \left
  (dr_J^2+\sinh^2 r_J d\Omega^2\right)\right].
\end{equation}

For later convenience, we introduce several symbols that describe
background geometrical quantities. 
The expressions given here are valid in $R$ and $L$ but
the extension to other regions is straightforward. 
We denote the unit vectors in $t$-direction and in 
$r$-direction as $\xi^{\mu}$ and $n^{\mu}$, respectively. 
Then the background metric can be decomposed as 
\begin{eqnarray}
 g_{\mu\nu} & = & -\xi_{\mu} \xi_{\nu}+\gamma_{\mu\nu}
\cr
 & = & -\xi_{\mu} \xi_{\nu}+n_{\mu} n_{\nu}+\sigma_{\mu\nu},
\end{eqnarray}
where $\gamma_{\mu\nu}$ and $\sigma_{\mu\nu}$ are 
the metric of the $t=\,$const. hypersurface and 
that of the $t,r=\,$const. surface, respectively. 
Further they are related to the metric of a unit 3-hyperboloid 
$\hat\gamma_{\mu\nu}$ and that of a unit 2-sphere 
$\hat\sigma_{\mu\nu}$ by
\begin{equation}
 \gamma_{\mu\nu} = a^2 \hat \gamma_{\mu\nu},
\quad
 \sigma_{\mu\nu} = a^2 \sinh^2 r\, \hat \sigma_{\mu\nu}.
\end{equation}
The small Latin indices such as 
$i$ and $j$ represent the projection; 
$f_{i}:=\gamma_{i}{}^{\mu} f_{\mu}$ and  
the vertical bar $|$ denotes the covariant derivative 
with respect to $\hat\gamma_{ij}$. 
The capital Latin indices such as 
$A$ and $B$ represent the projection; 
$f_{A}:=\sigma_{A}{}^{\mu} f_{\mu}$ and  
the double vertical bar $||$ denotes the covariant derivative 
with respect to $\hat\sigma_{AB}$. 

\section{Gravitational Wave Perturbations}
We consider quantized gravitational wave perturbations on the 
background given in the preceding section. 
In general, the perturbed metric and 
the perturbed scalar field are given as
\begin{equation}
 \tilde g_{\mu\nu}=g_{\mu\nu}+h_{\mu\nu}
 =g_{\mu\nu}+a^2~ H_{\mu\nu},
\quad
 \tilde\sigma=\sigma+\delta\sigma, 
\end{equation}
but here we ignore the perturbation of the scalar field and set 
$\delta\sigma=0$. 
Strictly speaking, one has to go to the region $C$, where a Cauchy
surface exists, to quantize a field. For the metric perturbation,
this is done in Appendix E for completeness.
However, here we take a more intuitive approach by
considering the metric perturbation in the region $R$ or $L$ first.
For the present problem, it turns out this approach is as good as the 
rigorous one given in Appendix E.

In $R$ or $L$, we can impose the transverse traceless synchronous 
gauge condition on the metric perturbation. 
To make our statement explicit, 
we introduce the tensor harmonics 
on the 3-hyperboloid, $Y^{p\, lm}{}_{\mu\nu}=Y^{(+)p\, lm}{}_{\mu\nu}$ 
and $Y^{(-)p\, lm}{}_{\mu\nu}$, by 
\cite{Tom} 
\begin{eqnarray}
&&Y^{(\pm)p\, lm}{}_{\mu\nu} \,\xi^{\nu}=0,
\nonumber\\
\nonumber\\
&&Y^{(+)p\, lm}{}_{ij}=
  \sqrt{(l-1)l(l+1)(l+2)\Gamma(ip+l+1)\Gamma(-ip+l+1)\over 2
  p^2(p^2+1)\Gamma(ip)\Gamma(-ip)}
 {\cal G}^{(+)p\, lm}_{ij},
\nonumber\\
\nonumber\\
&&Y^{(-)p\, lm}{}_{ij}=
  \sqrt{(l-1)(l+2)\Gamma(ip+l+1)\Gamma(-ip+l+1)\over 
   2 l(l+1)(p^2+1)\Gamma(ip)\Gamma(-ip)}
 {\cal G}^{(-)p\, lm}_{ij},
\label{tenharm}
\end{eqnarray}
where
\begin{eqnarray}
&&{\cal G}^{(+)p\, lm}_{rr}  = {\cal T}^{pl}_1 Y\,,
\quad
{\cal G}^{(+)p\, lm}_{r A}= {\cal T}^{pl}_2
 Y_{||A}\,,
\nonumber \\
&&{\cal G}^{(+)p\, lm}_{AB}
= {\cal T}^{pl}_3
 Y_{||AB}
+ {\cal T}^{pl}_4
 Y {\hat\sigma_{AB}}\,,
\label{calGp}
\\
\nonumber\\
 && {\cal G}^{(-)}_{rr}=0,
\quad
 {\cal G}^{(-)}_{r A} = {\cal T}_5^{pl} {\cal Y}_{A},
\quad
 {\cal G}^{(-)}_{AB} = 2 {\cal T}_6^{pl} {\cal Y}_{AB},
\label{calGm}
\end{eqnarray}
and $Y:=Y_{lm}(\Omega)$ are the 2-dimensional spherical harmonics and 
\begin{equation}
{\cal Y}_A :=Y_{||C}~\hat\epsilon^C_{~A}, 
\quad 
{\cal Y}_{AB} :=Y_{||C(A}~\hat\epsilon^{C}_{~B)}
 =\ell(\ell+1)Y_{C(A}~\hat\epsilon^{C}_{~B)}\,.
\label{calY}
\end{equation}
The superscripts $(+)$ and $(-)$ denote parities of the harmonics
and $\hat\epsilon_{AB}$ is the unit anti-symmetric tensor
on the unit 2-sphere ($\hat\epsilon_{\theta\varphi}=\sin\theta$ etc.)
and $\hat\epsilon^A{}_B=\hat\sigma^{AC}\hat\epsilon_{CB}$.

The $r$-dependent parts of the tensor harmonics are expressed 
in terms of the function,
\begin{equation}
{\cal P}_{p\,l}(r)
:=
{P_{ip-{1\over 2}}^{-l-{1\over 2}}(\cosh r)\over\sqrt{\sinh r}},
\label{fpldef}
\end{equation}
as 
\begin{eqnarray}
{\cal T}^{pl}_1 & = & {1\over \sinh^2 r  }{\cal P}_{pl}(r )\,,
\nonumber \\
{\cal T}^{pl}_2 & = & {1\over l(l+1)}
\left(\partial_{r }+{\coth r }\right){\cal P}_{pl}(r )\,,
\nonumber \\
{\cal T}^{pl}_3 & = & 
{2\sinh^2 r  \over (l-1)l(l+1)(l+2)}
\left(\coth r \partial_{r }
-\left\{p^2-1-{l(l+1)+2\over 2 \sinh^2 r }
\right\}\right){\cal P}_{pl}(r )\,,
\nonumber \\
{\cal T}^{pl}_4 & = & {\sinh^2 r  \over (l-1)(l+2)}
\left(\coth r \partial_{r }
-\left\{p^2-1-{2\over \sinh^2 r }
\right\}\right){\cal P}_{pl}(r )\,,
\cr
{\cal T}_5^{pl} & = & {\cal P}_{pl}(r ), 
\cr
{\cal T}_6^{pl} & = & 
   {\sinh^2 r\over (l-1)(l+2)}\left(\partial_{r } +2\coth\, 
    r \right){\cal P}_{pl}(r )\,,
\label{calT}
\end{eqnarray}
where $P^\mu_\nu(z)$ is the associated Legendre function of the first
kind.
These harmonics satisfy spatial transverse traceless 
gauge condition:
\begin{equation}
 Y^{p\, lm}{}_{ij}{}^{|j}=0, \quad Y^{p\, lm}{}_{i}{}^{i}=0,
\end{equation}
and have the eigenvalue $p^2+3$ as
\begin{equation}
 \left[{}^{(3)}\triangle+(p^2+3)\right]Y^{p\, lm}{}_{ij}
 =0, 
\label{tLap}
\end{equation}
where ${}^{(3)}\triangle$ is the tensor Laplacian operator on the unit
3-hyperboloid. 
Note that, by construction, there are no $l=0$ and $1$ modes in these
harmonics.
For positive $p^2$ modes, they are normalized as 
\begin{equation}
\int d\Sigma~
\hat\gamma^{ii'} \hat\gamma^{jj'}
 Y^{p\, lm}{}_{ij}\overline{Y^{p'\, l'm'}{}_{i'j'}}
  =\delta(p-p')\delta_{l,l'}\delta_{m,m'}\,,
\end{equation}
where $d\Sigma=\sinh^2r\,drd\Omega$ is the surface element on the unit
3-hyperboloid. 
For negative $p^2$ modes, they are not normalizable. 
Thus it might be inappropriate to call them 
harmonics but we do so here and  
define them by Eq.~(\ref{tenharm}).

Using the harmonics, we expand the metric 
perturbations as 
\begin{equation}
 H_{\mu\nu}=H_{(+)\mu\nu}+H_{(-)\mu\nu},
\end{equation}
and 
\begin{eqnarray}
 H_{(+)\mu\nu}=\sum_{p,l,m} 
 U^{(+)}_{p\, lm}(t) Y^{(+)p\, lm}_{\mu\nu}(r ,\Omega),
\cr
 H_{(-)\mu\nu}=\sum_{p,l,m} 
 U^{(-)}_{p\, lm}(t) Y^{(-)p\, lm}_{\mu\nu}(r ,\Omega).
\end{eqnarray}
Then the perturbed Einstein equation 
in $C$ becomes
\begin{equation}
 \left[{1\over a^3(T)}{d\over dT} a^3(T)
   {d\over dT} + {p^2+1\over a^2(T)}\right] 
   {U_{p\, lm}(T)}=0.
\label{Ueq}
\end{equation}
The equations in the other regions are obtained by 
the analytic continuation. 

This equation is exactly the same one that was discussed
in Ref.~\citen{ST96} for a noninteracting 
massless scalar field without
coupling to the metric perturbation. 
In the scalar case, we showed that 
there is one supercurvature mode ($p^2<0$) for each $l,m$ 
independent of the detail of the model
under the thin wall approximation. 
With this result in mind, in the subsequent two sections, 
we discuss subcurvature modes ($p^2>0$) and 
supercurvature modes separately.  

\subsection{subcurvature modes}
{}For subcurvature modes, the correspondence between 
the cases of the massless scalar perturbation
and the gravitational wave perturbation 
is exact. 

When $p^2>0$ 
the $r$-dependent parts of the harmonics 
vanish fast enough as $r\rightarrow \infty$. 
Therefore we can choose 
the union of a $t_L=\,$const. 
and a $t_R=\,$const. hypersurfaces as a surface on which 
the canonical commutation relations are set, 
although it is not a Cauchy surface.\footnote
{See the discussion given above Eq.~(2.15) of Ref.~\citen{STY95}.}

As given in Appendix A, 
in $L$ and $R$,
foliating the spacetime by $t_J=\,$const. hypersurfaces, 
the 2nd variation of the action becomes
\begin{equation}
\delta^{(2)}{\cal L}=
 {1\over 8\kappa}\sum_{\sigma=\pm}\sum_{J=R,L}
  \int_{-\infty}^{+\infty} dp \sum_{l,m} 
 \int d\tau_J\, a(\tau_{J})^3 
 \left(\left\vert
  {dU^{(\sigma)}_{p\, lm} \over d\tau_J} 
  \right\vert^2
  -{(p^2+1)\over a(\tau_J)^2} \vert U^{(\sigma)}_{p\, lm}\vert^2
 \right),
\label{2action}
\end{equation}
where $\tau_J=t_J/H_J$.
This action, for each definite parity, is the same as that 
of the massless scalar field $\phi$ 
with the decomposition,  
\begin{equation}
 \phi={1\over \sqrt{4\kappa}}\sum_{p,l,m} U_{p\, lm}(\tau)
 Y^{p\, lm}(r,\Omega), 
\end{equation}
where $Y^{p\, lm}(r,\Omega)$ is the normalized 
scalar harmonics on a unit 3-hyperboloid. 
The only difference is the presence of the $l=0$, $1$ modes
in the scalar case.
Then if we assume that the positive frequency functions 
of gravitational wave perturbations are determined by the 
same analyticity as in the case of the scalar field, 
the results obtained in Ref.~\citen{ST96} can be reinterpreted 
for the present problem.\footnote{
Strictly speaking, we have not proven 
that the prescription taken in Ref.~\citen{ST96} to 
determine the positive frequency functions for
the quantum state after bubble nucleation 
is also applicable to the present problem. 
The discussion to justify the prescription was given 
for perturbations of a scalar field in
Ref.~\citen{TS94} but 
the discussion in it did not 
take care of the case with gauge degrees of freedom. 
Thus here we adopt the analogy to the scalar case 
just by assumption.}  
Since the final expression for 
the amplitude of fluctuations of the 
scalar field, $\delta\phi$, 
was written in terms of the curvature perturbation ${\cal R}$ 
in Ref.~\citen{ST96},
it may be helpful to show here 
the relation; 
\begin{equation}
 {\cal R} = - {H\over \dot\phi(t)}\delta\phi.
\end{equation}
Then Eq.~(23) of Ref.~\citen{ST96} is reinterpreted to give 
the amplitude of the gravitational wave perturbations 
in the thin wall approximation: 
\begin{equation}
 \langle U_{p\, lm}^2\rangle ={4\kappa H^2\coth \pi p\over 2p(1+p^2)}
 (1-y),
\label{spec}
\end{equation}
where
\begin{equation}
1-y=
 1-{(\Delta s)^2 \cos\tilde p
     +2p \Delta s\sin \tilde p
      \over \cosh \pi p\,(4p^2+(\Delta s)^2)}\raisebox{-5pt}{,}
\end{equation}
with $\Delta s:=\sin t_{C,R}\vert_{wall} -\sin t_{C,L}\vert_{wall}$ 
and 
\begin{equation}
 \tilde p=p\ln\left(\displaystyle
{1+\sin t_{C,R}\vert_{wall}\over 1-\sin t_{C,R}\vert_{wall}}\right)
 \raisebox{-5pt}{.}
\end{equation}
It is important to note that $1-y\propto p^2$ as $p\to0$.
Although we have adopted the thin wall approximation here,
it is straightforward to extend the present analysis to the general case
and our conclusion that $1-y\propto p^2$ as $p\to0$ remains true.

As first pointed out by Allen and Caldwell,\cite{AlCal}
 if $y=0$, which would be the
case for the pure de Sitter background, the even parity gravitational
wave spectrum would have an infrared divergence in the limit
$p\to0$ because of the extra factor of $1/p^2$ in the normalization
factor of the tensor harmonics (see Eq.~(\ref{tenharm})).
 However, as soon as the effect of the presence of the bubble is
taken into account, this divergence disappears.
Thus for a realistic model of one-bubble open inflation, the
gravitational wave spectrum shows no pathological feature.

\subsection{supercurvature modes}
A supercurvature mode is a solution 
of Eq.~(\ref{Ueq}) regular at both boundaries, $T=T_R$ and $T_L$,
which may exist discretely at $p^2\leq0$.\footnote{Here
we use the terminology `supercurvature' in a broader sense to include
the case of $p^2=0$.}
Let us put aside the case of $p^2=0$ for a while. 
Then, as mentioned before,
it was shown in Ref.~\citen{ST96} under the thin wall approximation
that there 
is only one supercurvature mode for each $l$ and $m$. 
It is a trivial solution $U(T)=\,$const. for $p^2=-1$. 
However, different from the massless scalar case, 
this supercurvature mode is unphysical in the 
present case because it turns out to be 
just a gauge degree of freedom as shown in Appendix B. 
Thus in the thin wall case, there is no supercurvature mode 
in the gravitational wave perturbations. 
In Appendix C, we prove the absence of the 
supercurvature modes in general without the thin wall approximation.

However, one might worry if the absence of 
supercurvature modes would depend on our choice of gauge.  
That is, a metric perturbation described by a singular solution of 
Eq.~(\ref{Ueq}) could be transformed to a regular metric perturbation by
a different choice of gauge. For $p^2<0$ but $p^2\neq-1$, 
using the mutual independence among the scalar, vector and tensor
harmonics, it can be
shown without trouble that such a gauge transformation do not exist. 

For even parity $p^2=0$ modes, however, there is a problem that
they become degenerate with scalar perturbation modes with 
$p^2=-4$.\cite{HAMA,Garriga,YST96} 
Therefore, we must consider the scalar perturbation at the same time when
discussing regularity of the metric. In Appendix D, we 
treat this problem and show that there exists no gauge transformation
that makes the metric regular for even parity $p^2=0$ modes.
Thus it is concluded that there exists no supercurvature modes for
gravitational wave perturbations.

\section{Summary and discussion}
{}In this paper we have derived the spectrum of gravitational 
wave perturbations in the context of the open inflationary 
universe scenario. 
We have assumed that the quantum state after bubble nucleation 
is given by ``the Euclidean vacuum state'' that is 
determined by the analyticity of modes when they are 
continued to the Euclidean region.
Under this assumption and in the thin wall approximation,
we have explicitly obtained the spectrum (\ref{spec}).
An important feature of the spectrum is that it is infrared finite as
opposed to the case of pure de Sitter background.
We have also found that there is no discrete spectrum that comes from
supercurvature modes.
At a glance, there seemed to exist supercurvature modes 
at $p^2=-1$ but it is shown to be an illusion due to gauge 
degrees of freedom. 

A subtlety associated with the even parity $p^2=0$ modes that they
become degenerate with the $p^2=-4$ scalar perturbation modes
has been also resolved.
Taking account of all the degrees of freedom of metric perturbations,
we have shown that these modes do not exist.
In the previous analyses\cite{HAMA,Garriga,YST96} 
without taking into account the gravitational
degrees of freedom,
the scalar $p^2=-4$ modes occupied a special position because 
they existed independent of the detail of the potential 
of tunneling scalar field and were called 
wall fluctuation modes. 
Our result implies that these modes cease to exist once 
the gravitational degrees of freedom are taken into account.
This result is consistent with that obtained by 
Kodama et al.\cite{KodIsh} in the case of infinitely thin domain wall 
with vanishing potential energy in both vacua. 
We should note, however, that our result does not exclude the
possibility that a discrete mode describing the wall fluctuation exists
with a shifted eigenvalue other than $p^2=-4$.

\section*{Acknowledgments}
We thank Y. Mino, A. Ishibashi and H. Ishihara for helpful discussions. 
This work was supported in part by Monbusho Grant-in-Aid for
Scientific Research No.07304033.
\appendix

\section{2nd variation of the action}

In this Appendix we derive 2nd variation of the action given 
in Eq.~(\ref{2action}). Here we omit the subscript $J$ 
for simplicity. 

Since we are interested only in the 2nd order variation, 
we compute the terms quadratic in $h_{\mu\nu}$ in the 
Einstein-Hilbert action:
\begin{eqnarray}
\delta^2(\sqrt{-g}R)
&=&(\delta^2 \sqrt{-g}g^{\mu\nu})R_{\mu\nu}
+(\delta \sqrt{-g} g^{\mu\nu})\delta R_{\mu\nu}
+\sqrt{-g}g^{\mu\nu}\delta^2 R_{\mu\nu}
\nonumber\\
&=&
\sqrt{-g}\left(h^{\mu\rho}h_{\rho}{}^{\nu}-{1\over2}h h^{\mu\nu}
   +\left({1\over 8}h^2 
   -{1\over 4}h_{\rho\sigma}h^{\rho\sigma}\right)g^{\mu\nu}\right)
R_{\mu\nu}
\cr &&\quad
-\sqrt{-g}(h^{\mu\nu}-{1\over2}h g^{\mu\nu})
\delta R_{\mu\nu}
+\sqrt{-g}g^{\mu\nu}\delta^2 R_{\mu\nu},
\label{2ndvar}
\end{eqnarray}
Now
\begin{eqnarray}
\delta R_{\mu\nu}
&=&(\delta \Gamma^\rho{}_{\mu\nu})_{;\rho}
-(\delta \Gamma^\rho{}_{\mu\rho})_{;\nu}\,,
\\
\delta^2 R_{\mu\nu}
&=&(\delta^2 \Gamma^\rho{}_{\mu\nu})_{;\rho}
-(\delta^2 \Gamma^\rho{}_{\mu\rho})_{;\nu}
+\delta \Gamma^\sigma{}_{\rho\sigma}
\delta \Gamma^\rho{}_{\mu\nu}
-\delta \Gamma^\sigma{}_{\rho\nu}
\delta \Gamma^\rho{}_{\mu\sigma}\,.
\end{eqnarray}
Inserting these into Eq.~(\ref{2ndvar}) and using
\begin{eqnarray}
\delta \Gamma^\rho{}_{\mu\nu}={1\over2}
(h^\rho{}_{\mu; \nu}+h^\rho{}_{\nu; \mu}-h_{\mu\nu}{}^{;\rho}),
\end{eqnarray}
we obtain the 2nd order variation of the Einstein-Hilbert action,
\begin{eqnarray}
L_G^{(2)}
&=&{1\over 2\kappa}{1\over\sqrt{-g}}\delta^2 (\sqrt{-g}R)
\nonumber\\
&&\quad=L_{R}^{(2)}+{1\over 8\kappa}\left[
-h_{\mu\nu;\rho}h^{\mu\nu;\rho}+2h_{\mu\nu;\rho}h^{\rho\mu; \nu}
-2h_{\mu\nu}{}^{;\nu}h^{;\mu}+h_{;\mu}h^{;\mu}\right], 
\label{LagG}
\end{eqnarray}
where
\begin{equation}
L_{R}^{(2)}=
{1\over 8\kappa}[
 \left(4h^{\mu\rho}h_{\rho}{}^{\nu}-{2}h h^{\mu\nu}\right)
 R_{\mu\nu}
 +\left({1\over 2}h^2 
   -h_{\rho\sigma}h^{\rho\sigma}\right)R.
\end{equation}
Noting that 
\begin{eqnarray}
 R & = & 6\left({\ddot a\over a}
    +\left({\dot a\over a}\right)^2
    -{1\over a^2}\right),
\cr
 R_{\mu\nu} & = & 
    g_{\mu\nu}\left({\ddot a\over a}
    +2\left({\dot a\over a}\right)^2
    -{2\over a^2}\right)-2 \xi_{\mu} \xi_{\nu} 
    \left({\ddot a\over a}
    -\left({\dot a\over a}\right)^2
    +{1\over a^2}\right),
\end{eqnarray}
where $\xi^\mu$ is the unit vector normal to the $t=\,$constant
hypersurface,
we obtain
\begin{eqnarray}
 L_{R}^{(2)}=&&{1\over 8\kappa}\biggl[
 4\left(2 h^{\mu\rho} h_{\rho}{}^{\nu}- h h^{\mu\nu}\right)
 \xi_{\mu} \xi_{\nu} \left(-{\ddot a\over a}
    -\left({\dot a\over a}\right)^2+{1\over a^2}\right)
\cr
 &&+\left(h^{2} - 2 h_{\mu\nu} h^{\mu\nu}\right)
 \left({\ddot a\over a}
    -\left({\dot a\over a}\right)^2+{1\over a^2}\right)\biggr],
\end{eqnarray}
where the dot denote the derivative with respect to the cosmological
time, $t/H$.

We also need the matter part of the Lagrangian:
\begin{eqnarray} 
L_m^{(2)} & = & {1\over\sqrt{-g}}\delta^2 
 \left(\sqrt{-g} \left(
  -{1\over 2}g^{\mu\nu}\xi_{\mu}\xi_{\nu}
    \dot\sigma^2
  -V(\sigma)\right)\right),
\cr
 & = & -{1\over 2}\left(2 h^{\mu\rho} h_{\rho}{}^{\nu}- 
   h h^{\mu\nu}\right) \xi_{\mu} \xi_{\nu} \dot\sigma^2+
  {1\over 8}\left(h^{2} - 2 h_{\mu\nu} h^{\mu\nu}\right)
  \left({1\over 2}\dot\sigma^2 -V(\sigma)\right).
\end{eqnarray}

Then, using the equations of motion,
\begin{eqnarray}
&& \left({\dot a\over a}\right)^2
    -{1\over a^2}
  ={\kappa\over 3} 
   \left({1\over 2}\dot\sigma^2+V(\sigma)\right),
\cr
&&{\ddot a\over a}
  =-{\kappa\over 3}\left(
   \dot\sigma^2-V(\sigma)\right),
\end{eqnarray}
the summation of $L_{R}^{(2)}$ and $L_{m}^{(2)}$ 
simplifies to
\begin{equation}
L_{R}^{(2)} + L_{m}^{(2)}=
 -{1\over 8\kappa}\left(h^{2} - 2 h_{\mu\nu} h^{\mu\nu}\right)
 \left({\ddot a\over a}
    +2\left({\dot a\over a}\right)^2-{2\over a^2}\right).
\label{Lagadd}
\end{equation}

Now, taking into account the facts that $h_{\mu\nu} \xi^{\nu}=0$ 
and $h=0$, 
the 2nd variation of the Lagrangian 
reduces to
\begin{eqnarray}
 L^{(2)}
&&={1\over 8\kappa}\biggl[2 h_{ij} h^{ij}
 \left({\ddot a\over a}
    +2\left({\dot a\over a}\right)^2-{2\over a^2}\right)
\cr && \quad
+ \gamma^{\mu\mu'} \gamma^{\nu\nu'}
\bigl[\xi^{\rho}\xi^{\rho'}\left(
  h_{\mu\nu;\rho}h_{\mu'\nu';\rho'}
  -4 h_{\mu\nu;\rho}h_{\rho'\mu';\nu'}\right)
\cr &&\quad\quad
  + \gamma^{\rho\rho'}\left(
  -h_{\mu\nu;\rho}h_{\mu'\nu';\rho'}
  +2 h_{\mu\nu;\rho}h_{\rho'\mu';\nu'}\right)
\biggr].
\end{eqnarray}
The nonvanishing components of $h_{\mu\nu;\rho}$ 
are explicitly written as 
\begin{eqnarray}
 h_{\mu\nu;\rho}~ 
 \xi^{\mu}  \gamma^{\nu}{}_{i} \gamma^{\rho}{}_{j}
 & = & - {\dot a\over a} h_{ij},
\cr
  h_{\mu\nu;\rho}~ 
 \gamma^{\mu}{}_{i} \gamma^{\nu}{}_{j} \xi^{\rho} 
 & = &\dot h_{ij} - 2{\dot a\over a}h_{ij},
\cr
  h_{\mu\nu;\rho}~ 
 \gamma^{\mu}{}_{i} \gamma^{\nu}{}_{j} 
  \gamma^{\rho}{}_{j} 
 & = & h_{ij|k}.
\end{eqnarray}

Then using formulas of the $\Sigma$-integration of the harmonics 
\begin{eqnarray} 
 \int d\Sigma~{\hat \gamma^{ii'}\hat \gamma^{jj'}\hat \gamma^{kk'}}
   Y^{p\, lm}{}_{ij|k}\overline{Y^{p'\, lm}{}_{i'j'|k'}}
 &=& (p^2+3)\delta(p-p'),
\cr
 \int d\Sigma~{\hat \gamma^{ii'}\hat \gamma^{jj'}\hat \gamma^{kk'}}
   Y^{p\, lm}{}_{ij|k}\overline{Y^{p'\, lm}{}_{k'i'|j'}}
 &=& 3 \delta(p-p') ,
\end{eqnarray}
we obtain the resulting formula (\ref{2action}). 

\section{$p^2=-1$ mode}

Looking at Eq.~(\ref{Ueq}), it is easy to find 
that there is a trivial supercurvature 
mode solution at $p^2=-1$, which is given by $U(T)=\,$const.. 
Thus the corresponding metric perturbation is given by 
\begin{equation}
  h_{\mu\nu}\propto a^2 {\cal G}_{\mu\nu}, 
\label{hgauge}
\end{equation}
where ${\cal G}_{\mu\nu}$ are the unnormalized harmonics 
on the unit 3-hyperboloid 
defined by Eqs.~(\ref{calGp}) and (\ref{calGm}).
In this Appendix, we show that 
this mode is not a physical one 
but a gauge mode. 

To see this, we introduce the unnormalized 
vector harmonics on the unit 3-hyperboloid, ${\cal W}_i^{(\pm)p\, lm}$,
which are defined by 
\begin{eqnarray}
 {\cal W}^{(+)p\, lm}_r & = & {\cal V}_1^{pl} Y,
\quad
 {\cal W}^{(+)p\, lm}_A = {\cal V}_2^{pl} Y_{||A},
\cr
 {\cal W}^{(-)p\, lm}_r & = & 0,
\quad
 {\cal W}^{(-)p\, lm}_A = {\cal V}_3^{pl} {\cal Y}_{A},
\end{eqnarray}
where $Y=Y_{lm}(\Omega)$ and ${\cal Y}_A$ is defined in Eq.~(\ref{calY}). 
The $r$-dependent parts are 
\begin{equation}
 {\cal V}_1^{pl} = {1\over \sinh r}{\cal P},
\quad
 {\cal V}_2^{pl} = {1\over l(l+1)}\partial_r \left(\sinh r{\cal P}\right),
\quad
 {\cal V}_3^{pl} = {\sinh r}{\cal P},
\end{equation}
where $\cal P$ is defined in Eq.~(\ref{fpldef}).
The harmonics satisfy
\begin{eqnarray}
 &&\left({}^{(3)}\triangle+p^2+2\right) {\cal W}_i=0,
\label{B4}
\\
 && {\cal W}_{i}{}^{|i}=0,
\\
 && 2{\cal W}_{(i|j)}{}^{|j}=-(p^2+4){\cal W}_i,
\label{sijdiv}
\\
 &&\left({}^{(3)}\triangle+p^2+6\right) {\cal W}_{(i|j)}=0.
\label{sijlap}
\end{eqnarray}

{}From Eq.~(\ref{sijdiv}) 
one can see that the tensor constructed from 
the vector harmonics with $p_v^2=-4$ is transverse and traceless. 
Here we appended the subscript $v$ to $p$ to stress that it is the
eigenvalue of the vector harmonics defined by Eq.~(\ref{B4}). 
Thus ${\cal W}_{(i|j)} {(p_v^2=-4)}$ can be regarded as tensor
harmonics. Comparing Eq.~(\ref{sijlap}) with Eq.~(\ref{tLap}),
the corresponding eigenvalue $p_t$ of the tensor harmonics is found as
$p_t^2=p_v^2+3=-1$.
In fact, we can calculate ${\cal W}_{(i|j)}$ explicitly 
as 
\begin{eqnarray}
  {\cal W}_{(i|j)}\hat n^i \hat n^j &=& \partial_r {\cal W}_r,
\cr 
  {\cal W}_{(i|j)}\hat n^i\sigma^j{}_A &=& 
   {1\over 2}\left[\left(
    \partial_r -2\coth r\right){\cal W}_A +\partial_A{\cal W}\right],
\cr
  {\cal W}_{(i|j)}\sigma^i{}_A\sigma^j{}_B &=& 
    {\cal W}_{(A|B)} +\hat\sigma_{AB}\sinh r\cosh r {\cal W}_{r},
\end{eqnarray}
where $\hat n^i$ is the unit normal in the $r$-direction on the unit
3-hyperboloid and $\sigma_{ij}=\hat\gamma_{ij}-\hat n_i\hat n_j$.
Using the fact that 
\begin{equation}
{d\over dr}{{\cal P}{(p^2=-4)}\over \sinh r}=
 (l-1){{\cal P}{(p^2=-1)}\over \sinh^2 r}, 
\end{equation}
which follows from the derivative recursion relation of the associated
Legendre functions, it can be verified that 
\begin{eqnarray}
 {\cal W}^{(+)}_{(i|j)}{(p^2=-4)} & = & 
 (l-1) {\cal G}^{(+)}_{ij}{(p^2=-1)},
\cr
 {\cal W}^{(-)}_{(i|j)}{(p^2=-4)} & = & 
 {(l-1)\over 2} {\cal G}^{(-)}_{ij}{(p^2=-1)}. 
\end{eqnarray}

Now we consider a purely spatial gauge transformation 
(infinitesimal coordinate transformation),
\begin{equation}
 x^{\mu}\rightarrow x^{\mu}+\zeta^{\mu},
\end{equation}
with $\zeta_{\mu}\xi^{\mu}=0$ and 
$\zeta_{\mu}\propto {\cal W}_{\mu}{(p^2=-4)}$. 
It gives the change of the metric;
\begin{eqnarray}
&& 2\zeta_{(\mu;\nu)}\xi^{\nu}=0,
\cr
&& 2\zeta_{(i;j)}\propto a^2 {\cal G}_{ij}(p^2=-1),
\end{eqnarray}
without disturbing the value of the scalar field.
This is just the metric perturbation 
for the $p^2=-1$ mode given in Eq.~(\ref{hgauge}). 

\section{absence of supercurvature modes}
Here we show that there exist no supercurvature modes 
in gravitational wave perturbations without using the 
thin wall approximation. 

A supercurvature mode is a mode corresponding 
to a regular solution of Eq.~(\ref{Ueq}) with $p^2\leq0$.
The problem to find a supercurvature mode is analogous to 
that to find a bound state in the quantum mechanics. 
Here, we note that a regular solution means the one that gives a regular
metric perturbation. An inspection of Eq.~(\ref{Ueq}) reveals that
the solution behaves as $a^{-1\pm\Lambda}$ ($\Lambda=\sqrt{-p^2}$) for
$p^2<0$ and as $a^{-1}$ or $a^{-1}\ln a$ for $p^2=0$ as $T\to T_J$. 
Then it can be shown that the metric perturbation is regular 
if the solution behaves as $a^{-1+\Lambda}$.
Thus a supercurvature mode is an eigen mode that satisfies the boundary
condition $U\to a^{-1+\Lambda}$ as $T\to T_J$.

{}First we introduce a new coordinate, $s$, defined by 
\begin{equation}
 dT=a^3 ds, \quad a(s=0)=a_{max},
\end{equation} 
where $a_{max}$ is the maximum value of $a$ in C. 
Then Eq.~(\ref{Ueq}) is rewritten as 
\begin{equation}
 \left[{d^2\over ds^2}
   + (p^2+1) a^6(s)\right]{U_{p\, lm}(s)}=0.
\label{Ueqmod}
\end{equation}
If there is a potential $Q(s)$ which satisfies 
\begin{equation}
  Q(s)\ge a^6(s)~~ {\rm for}~{}^\forall s\,,
\quad Q(s)>a^6(s)~~ {\rm for}~{}^\exists s\,, 
\label{Qcondi}
\end{equation}
and if the equation 
\begin{equation}
 \left[{d^2\over ds^2}
   +(p^2+1) Q(s)\right] X=0, 
\label{UeqmodQ}
\end{equation}
has $N$ supercurvature modes (bound state solutions)
with one of them at $p^2=0$, 
then the number of supercurvature modes 
of Eq.~(\ref{Ueqmod}) is less than $N$. 

Integrating the equations of motion (\ref{sigmaeq}), 
we obtain the following inequality:
\begin{eqnarray}
 {1\over 2}\dot\sigma^2-V(\sigma(T))
 &=& {1\over 2}(\dot\sigma\vert_{T=T_{max}})^2
    -V(\sigma(T_{max}))
  -3\int_{T_{max}}^T dT' 
     {\dot a\over a}\dot\sigma^2
\cr
 &\ge & {1\over 2}(\dot\sigma\vert_{T=T_{max}})^2
    -V(\sigma(T_{max}))=:-{3\over \kappa}\tilde H^2, 
\end{eqnarray}
where the dot denotes the derivative with respect to 
$T$. 
The equality holds only for $T=T_{max}$. 
Then from Eq.~(\ref{aeq}), we find 
\begin{equation}
 \left({\dot a\over a}\right)^2 -{1\over a^2}
 \ge -\tilde H^2.
\label{ineq}
\end{equation}
Note that from the fact that the equality holds 
at $T=T_{max}$, i.e., at $s=0$, where $\dot a=0$, 
one finds
\begin{equation}
a_{max}=\tilde H^{-1}. 
\end{equation}
We introduce the scale factor $\tilde a$ of the 
de Sitter space with the radius $\tilde H^{-1}$. 
Then $\tilde a$ satisfies 
\begin{equation}
 \left({\dot {\tilde a}\over \tilde a}\right)^2 
   -{1\over \tilde a^2}
 = -\tilde H^2.
\label{ineqp}
\end{equation}
Rewriting Eqs.~(\ref{ineq}) and (\ref{ineqp}) by using the
$s$-coordinate, we obtain 
\begin{equation}
 A\left((\partial_s A)^2-1\right)\ge
 B\left((\partial_s B)^2-1\right),
\label{keyeq}
\end{equation}
where $A=1/2a^2$ and $B=1/2\tilde a^2$ and the equality holds only at
$s=0$.

Now we prove that $A>B$, i.e., 
$a(s)<\tilde a(s)$ for $^{\forall} s$ except at $s=0$. 
Suppose $A\leq B$ at $s=s_1>0$. 
Then Eq.~(\ref{keyeq}) tells us that 
$(\partial_s A)^2>(\partial_s B)^2$ 
because $A$ and $B$ are positive. 
Since $\partial_s A>0$ for $s>0$, 
this implies $\partial_s A>\partial_s B$. 
Therefore we would conclude that 
$B(s_1)-A(s_1)<B(s=0)-A(s=0)$. 
However, this contradicts with the fact 
$A(s=0)=B(s=0)=\tilde H^{2}/2$. 
Thus $A>B$ for $s>0$. 
A parallel discussion holds also in the case $s_1<0$. 
Thus we conclude that $A>B$ except at $s=0$. 

Now if we set $Q(s)=\tilde a^6(s)$, 
the condition (\ref{Qcondi}) is satisfied and 
Eq.~(\ref{UeqmodQ}) is equivalent to the 
equation to determine the mode function of a 
minimally coupled massless scalar field in 
the pure de Sitter space, which 
is known to have one supercurvature mode 
at $p^2=-1$ ($U=\,$const.) and one marginally regular mode at
$p^2=0$ ($U\propto \dot a/a$).
So Eq.~(\ref{Ueq}) has only one supercurvature mode, 
which is the one at $p^2=-1$. 
But it was shown in Appendix B that this unique supercurvature mode 
is an illusion due to gauge degrees of freedom. 
Thus we conclude that there is no supercurvature mode in 
gravitational wave perturbations. 

\section{$p^2=0$ modes}

Here we consider the even parity $p^2=0$ modes by taking full account of
the degeneracy between the scalar and tensor harmonics.
Since it is natural to require that a physical mode should give 
regular metric and scalar field perturbations, we examine whether
a regular solution exists in this degenerate case.

First we give the explicit relation between the scalar and tensor
harmonics when they are degenerate.
Let us introduce the unnormalized scalar harmonics on the unit
3-hyperboloid, ${\cal S}^{plm}={\cal P}_{pl}Y_{lm}$. They satisfy
\begin{eqnarray}
 &&\left({}^{(3)}\triangle+p^2+1\right){\cal S}=0,
\label{Qlap}
\\
 && {\cal S}_{ij}{}^{|j}=-{2\over3}(p^2+4){\cal S}_{|i}\,,
\label{Qijdiv}
\\
 &&\left({}^{(3)}\triangle+p^2+7\right){\cal S}_{ij}=0\,,
\label{Qijlap}
\end{eqnarray}
where 
\begin{equation}
{\cal S}_{ij}:={\cal S}_{|ij}+{p^2+1\over3}\hat\gamma_{ij}{\cal S}\,.
\label{Qijdef}
\end{equation}
One readily sees ${\cal S}_{ij}$ is transverse-traceless when $p^2=-4$.
Comparison of Eq.~(\ref{Qijlap}) with Eq.~(\ref{tLap})
immediately shows $p_t^2=p_s^2+4$, where $p_t$ and $p_s$ are the
eigenvalues of the tensor and scalar harmonics, respectively.
Hence the $p_s^2=-4$ modes are equivalent to tensor harmonics with
$p_t^2=0$. Further by construction, it is clear that they have even
parity. The explicit form of ${\cal S}_{ij}(p^2=-4)$ can be obtained by
noting the fact,
\begin{equation}
{d^2\over dr^2}{\cal P}(p^2=-4)={\cal P}(p^2=-4)+
l(l-1){{\cal P}(p^2=0)\over\sinh^2r}\,.
\end{equation}
Then we find
\begin{equation}
{\cal S}_{ij}(p^2=-4)=l(l-1){\cal G}^{(+)}_{ij}(p^2=0),
\end{equation}
where ${\cal G}^{(+)}_{ij}$ are given by Eq.~(\ref{calGp}).
Thus it is necessary (and sufficient) to consider scalar type
perturbations with $l\geq2$ 
when discussing the even parity $p_t^2=0$ modes.

We consider the scalar type perturbations in region $R$ or $L$ and
describe the perturbed metric as
\begin{equation}
d\tilde s^2=-(1+2A{\cal S})d\tau^2+2B{\cal S}_{|j}d\tau dx^j+
a(t)^2\left\{(1+2H_L{\cal S})\hat\gamma_{ij}
+2H_T{\cal S}_{ij}\right\}dx^idx^j\,,
\end{equation}
and the perturbed scalar field as
\begin{equation}
\tilde\sigma=\sigma+\varphi{\cal S}\,.
\end{equation}
Then the field equation for $\varphi$ is given by
\begin{eqnarray}
\ddot\varphi
&&+3{\dot a\over a}\dot\varphi+{p^2+1\over a^2}\varphi
+V''(\sigma)\varphi
\nonumber\\
&&-2A(\ddot\sigma+3{\dot a\over a}\dot\sigma)-\dot A\dot\sigma
+3\dot H_L\dot\sigma+{p^2+1\over a^2}B\dot\sigma=0.
\label{fieldeq}
\end{eqnarray}
The necessary components of the perturbed Einstein equations
are the $(0,0)$, $(0,i)$ and the traceless part of $(i,j)$ 
components. They are given, respectively, as
\begin{eqnarray}
&&2\left(3\left({\dot a\over a}\right)^2A-3{\dot a\over a}\dot H_L
-{\dot a\over a}\,{p^2+1\over a^2}B
-{p^2+4\over a^2}(H_L+{p^2+1\over3}H_T)\right)
\nonumber\\
&&\qquad\qquad
=\kappa(A\dot\sigma-\dot\sigma\varphi-V'\varphi)\,,
\label{00comp}\\
&&2\left({\dot a\over a}A-\dot H_L+{1\over a^2}B
-{p^2+4\over3}\dot H_T\right)=\kappa\dot\sigma\varphi\,,
\label{0icomp}\\
&&
{1\over a}(a^3\dot H_T)\dot{}\,-{1\over a}(aB)\dot{}\,
-A-H_L-{p^2+1\over3}H_T=0.
\label{ijcomp}
\end{eqnarray}

In order to solve the above set of equations, it is necessary to fix a
gauge. For this purpose, let us consider a gauge transformation induced
by an infinitesimal coordinate transformation,
\begin{equation}
\bar \tau=\tau+M{\cal S}\,,\quad \bar x^i=x^i+N{\cal S}^{|i}\,.
\end{equation}
Then the perturbation variables transform as
\begin{eqnarray}
&&\bar A=A-\dot M\,,
\nonumber\\
&&\bar B=B-a^2\dot N+M\,,
\nonumber\\
&&\bar H_L=H_L+{p^2+1\over3}N-{\dot a\over a}M\,,
\nonumber\\
&&\bar H_T=H_T-N\,,
\nonumber\\
&&\bar\varphi=\varphi-\dot\sigma M\,.
\label{E14}
\end{eqnarray}
Hence unless $\sigma$ is constant, which
is the case of pure de Sitter background, one can choose a gauge in
which $\varphi=H_L=0$ by using the above gauge degrees of freedom. 
Now specializing to the case of $p^2=-4$, one
finds from Eqs.~(\ref{00comp}) and (\ref{0icomp}) that
$A=0$ in this gauge. 
Then we find $B=0$ from the perturbed field equation~(\ref{fieldeq}).
Thus the only remaining variable is $H_T$. From Eq.~(\ref{ijcomp}),
we find it satisfies
\begin{equation}
{1\over a^3}{d\over d\tau}(a^3{d\over d\tau}H_T)+{1\over a^2}H_T=0\,.
\label{E15}
\end{equation}
Now going into the region $C$ by 
the identifications of $d\tau$ with $idT$ and $a$
with $ia$, we see that this equation exactly coincides with the one
for the tensor $p^2=0$ modes given by Eq.~(\ref{Ueq}).

Now let us examine the asymptotic behavior 
of the solution of Eq.~(\ref{E15}) at $a\rightarrow 0$ in $C$. 
Noting that $a\sim\vert T-T_J\vert=:\Delta T$ near the boundaries, 
$H_T$ have regular and singular 
solutions which behave as $\Delta T^{-1}$ and 
$\Delta T^{-1}\ln \Delta T$, respectively. 
Since the statement given in Appendix C 
holds also in the present case, any solution 
is singular either at $T_R$ (boundary to the region $R$)
or at $T_L$ (boundary to the region $L$). 
Thus we may conclude that there exists no discrete mode for $p^2=-4$
(i.e., no discrete even parity $p^2=0$ tensor mode) that would
contribute to the quantum fluctuations.

However, it is not yet completely clear if the singular behavior 
of $H_T$ is real. It may be absorbed by a gauge transformation.
Thus we have to show that there exists no gauge transformation that
makes the metric and scalar field perturbations regular.
The regularity can be examined by investigating the behavior 
of the perturbations as one approaches either of the two boundary light
cones.
Since the coordinates $T$ and $r_C$ are degenerate on the 
boundary light cone,
it is necessary to evaluate the metric components 
in non-degenerate coordinates, e.g.,
\begin{equation}
 \tilde t=\Delta T\cosh r_C\,,\quad \tilde r=\Delta T\sinh r_C.
\end{equation}
Then the components are related by 
\begin{equation}
 \left(
  \begin{array}{c}
  \Delta T^2 h_{TT} \\
  \Delta T h_{Tr} \\
  h_{rr}
 \end{array}\right)
=\left(
  \begin{array}{ccc}
  \tilde t^2 & 2 \tilde t\tilde r & \tilde r^2\\
  \tilde t\tilde r& \tilde t^2+\tilde r^2 & \tilde t\tilde r\\
  \tilde r^2 & 2 \tilde t\tilde r & \tilde t^2
 \end{array}\right)
\left(
  \begin{array}{c}
  h_{\tilde t\tilde t} \\
  h_{\tilde t\tilde r} \\
  h_{\tilde r\tilde r}
 \end{array}\right).
\end{equation}
When we take the limit to the boundary light cone, 
$\Delta T$ goes to $0$ while $\tilde t$ and $\tilde r$ 
stay finite.  
Thus it is required that 
the components,
$\Delta T^2 h_{TT}\sim \Delta T^2 A {\cal S}$ and 
$\Delta T h_{Tr}\sim \Delta T B \partial_{r_C}{\cal S}$,
should be regular.
Since the radial part of ${\cal S}$ 
behaves as $\exp{(\sqrt{-p^2}-1)r_C}=\exp{r_C} \sim \Delta T^{-1}$, 
$\Delta T A$ and $B$ should be regular on the boundary light cone.

Now we know that it is sufficient to consider a solution that behaves as
$H_T=\Delta T^{-1}\ln \Delta T$ at, say, the left boundary
in $\varphi=A=B=H_L=0$ gauge. 
The gauge transformation (\ref{E14}) is still valid 
in region $C$ with replacements, $M\rightarrow iM$ and 
$B\rightarrow iB$. 
Thus we must set $N= \Delta T^{-1}\ln \Delta T$ 
in order to remove the singular behavior of $H_T$. 
Then $B$ becomes singular. 
To remove the singular behavior of $B$, 
we should take $M=-\ln \Delta T$. 
Then, however, $\varphi$ becomes singular as 
$-\dot\sigma \ln \Delta T\sim \Delta T\ln \Delta T$. 
The perturbation of scalar field itself is finite but the 
derivative diverges. 
Thus we finally conclude that no regular (hence physical) discrete 
mode exists.

\section{Canonically reduced action for gravitational 
wave perturbations}
Here, we discuss the reduction of the action 
for gravitational wave perturbations in an open inflationary universe. 
We consider the metric perturbation in the region $C$, where a Cauchy
surface exists,
and take a canonical approach to reduce the degrees of 
freedom of the constrained system to the physical degrees of freedom. 
The discussion goes parallel to the case 
of gravitational waves in the Rindler universe given in 
Appendix A of Paper I.
The case of the Rindler universe, in which the scale factor is set 
to $a(T)=T$, is one special example of general cases but 
almost all the equations which appeared in the Rindler 
case hold with replacements:
\begin{eqnarray}
 &&\chi\rightarrow r,
\quad \partial_{\chi}\rightarrow \partial_r,
\quad \xi\rightarrow a(T),
\quad \partial_{\xi}-{m\over \xi}\rightarrow
   \partial_{T}-m{\dot a\over a},\quad (m=0,1,2,\cdots),
\cr
 && (e)\rightarrow (+), \quad (o)\rightarrow (-),
\label{replrule}
\end{eqnarray}
where the dot $~\dot{~}~$ represents the derivative with 
respect to $T$. 
Thus we only show here the necessary changes other than these
replacements.
In this Appendix the subscript $C$ in $r_C$ is suppressed
for notational simplicity. 

The Lagrangian for gravitational wave perturbations is given in 
Eqs.~(\ref{Lagadd}) and (\ref{LagG}). After analytic continuation to 
the region $C$, it becomes 
\begin{eqnarray}
L^{(2)}= &&{1\over 8\kappa}\left(h^{2} - 2 h_{\mu\nu} h^{\mu\nu}\right)
   \left({\ddot a\over a}
   +2\left({\dot a\over a}\right)^2-{2\over a^2}\right)
\cr&&
   +{1\over 8\kappa}\left(-h_{\mu\nu;\rho}h^{\mu\nu;\rho} 
   + 2 h_{\mu\nu;\rho}h^{\rho\mu;\nu}
   - 2 h_{\mu\nu}{}^{;\nu} h^{;\mu} + h_{;\mu} h^{;\mu}\right).
\label{A1}
\end{eqnarray}

We adopt the convention to denote the projection of tensors as 
\begin{eqnarray}
 f_{\xi} & := & f_{\mu}\, \xi^{\mu},
\cr
 f_{n} & := & f_{\mu}\, n^{\mu}=a^{-1} f_{r}. 
\end{eqnarray}
The relations,
\begin{eqnarray}
 n^{\mu}{}_{;\nu}& = &-{\dot a\over a} \xi^{\mu}n_{\nu}
                   +{\tanh r \over a} \sigma^{\mu}{}_{\nu}\,,
\cr
 \xi^{\mu}{}_{;\nu}& = &-{\dot a\over a} n^{\mu}n_{\nu}
                   +{\dot a \over a} \sigma^{\mu}{}_{\nu}\,,
\end{eqnarray}
are used in the calculations below.
Each component of covariant derivatives of the metric perturbation 
becomes 
\begin{eqnarray}
 h_{nn;n}
 & = & {1\over a} \partial_r h_{nn}-2{\dot a\over a}h_{n\xi},
\quad 
h_{nn;\xi}
  =  \partial_{T} h_{nn}, 
\quad h_{nn;A}
 = h_{nn||A}-{2\tanh r\over a}h_{n A},
\cr
 h_{n\xi;n}
 & = & {1\over a} \partial_r h_{n\xi}
    -{1\over a}\left(h_{\xi\xi}+h_{nn}\right),
\quad h_{n\xi;\xi}
 = \partial_{T} h_{n\xi}, 
\cr
 h_{n\xi;A}
 & = & h_{n\xi||A}-{\tanh r\over a}h_{\xi A}
                 -{\dot a\over a}h_{n A},
\quad h_{\xi\xi;n}
  = {1\over a} \partial_ r h_{\xi\xi}-2{\dot a\over a}h_{n\xi},
\cr
 h_{\xi\xi;\xi}
 & = & \partial_{T} h_{\xi\xi}, 
\quad h_{\xi\xi;A}
 = h_{\xi\xi ||A}-2{\dot a\over a}h_{\xi A},
\cr
\quad h_{nA;n}
 & = & {1\over a}\left(\partial_{r}-\tanh r\right) h_{nA}
       -{\dot a\over a}h_{\xi A},
\quad h_{nA;\xi}
 = \left(\partial_{T}-{\dot a\over a}\right) h_{nA}, 
\cr
 h_{nA;B}
 & = & h_{nA||B}+\left({\dot a\over a}h_{n\xi}-
  {\tanh r\over a}h_{nn}\right)\sigma_{AB}
   -{\tanh r\over a} h_{AB},
\cr
 h_{\xi A;n}
 & = & {1\over a}\left(\partial_{r}-\tanh r\right) h_{\xi A}
       -{\dot a\over a}h_{nA},
\quad h_{\xi A;\xi}
 = \left(\partial_{T}-{\dot a\over a}\right) h_{\xi A}, 
\cr
 h_{\xi A;B}
 & = & h_{\xi A||B}+\left({\dot a\over a}h_{\xi\xi}-
  {\tanh r\over a}h_{n\xi}\right)\sigma_{AB}
   -{\dot a\over a} h_{AB},
\cr
 h_{AB;n}
 & = & {1\over a}\left(\partial_{ r}-2\tanh r\right) h_{AB},
\quad h_{AB;\xi}
 = \left(\partial_{T}-2{\dot a\over a}\right) h_{AB}, 
\cr
 h_{AB;C}
 & = & h_{AB||C}+\left(2{\dot a\over a}h_{\xi (A}\sigma_{B)C}-
  {2\tanh r\over a}h_{n(A}\sigma_{B)C}\right),
\end{eqnarray}
where we used the abbreviated notation such as 
$h_{nA;\xi}\equiv h_{\mu\nu;\rho}n^{\mu}\sigma^{\nu}_{~A}\xi^{\rho}$. 
Below we expand the metric perturbation in terms of the spherical
harmonics and consider the even and odd parity modes separately.

\subsection{even parity}
Concentrating on the even parity modes, we expand the variables 
by using the spherical harmonics $Y=Y_{\ell m}(\Omega)$,
\begin{eqnarray}
&& h^{(+)}_{nn}=\sum H^{(+)\ell m}_{nn}Y, \quad 
   h^{(+)}_{n\xi}=\sum H^{(+)\ell m}_{n\xi}Y, \quad 
   h^{(+)}_{\xi\xi}=\sum H^{(+)\ell m}_{\xi\xi}Y, 
\cr
&& h^{(+)}_{nA}=\sum H^{(+)\ell m}_{n}Y_{||A},\quad 
   h^{(+)}_{\xi A}=\sum H^{(+)\ell m}_{\xi}Y_{||A},
\cr 
&& h^{(+)}_{AB}=\sum \left(w^{(+)\ell m} Y \hat\sigma_{AB}
                 +v^{(+)\ell m} Y^{(s)}_{AB}\right),
\label{vardef}
\end{eqnarray}
where 
\begin{equation}
 Y^{(s)}_{AB}={Y_{||AB}\over \ell(\ell+1)}+{1\over 2}\hat\sigma_{AB} Y. 
\end{equation}
Then the same argument that are given below Eq.~(A8) of 
Paper I holds with the replacements listed in 
Eq.~(\ref{replrule}) and the action can be reduced under 
the synchronous gauge condition. 
Finally we obtain the reduced action:
\begin{eqnarray}
 \int dr \int && dT~{\cal L}^{(+)}_{(red)}
  =  \sum_{\ell,m}{8\over (\ell-1)\ell(\ell+1)(\ell+2)}
    \int dr \int {dT\over a^3} 
 \Biggl[
    \overline{\Pi} \hat K(\hat K -1)\left(\partial_r w\right)
\cr && -{1\over 2}\left({1\over\cosh^2 r}\overline{\Pi}
   \hat K(\hat K -1)\Pi+
    \overline{w} \hat K(\hat K -1)
    \left\{\ell(\ell+1)+\hat K\cosh^2 r
    \right\} w\right)\Biggr],\cr &&
\label{redact}
\end{eqnarray}
where $\Pi$ is defined by 
\begin{equation}
 \Pi^{(+)\ell m}
 :=-\cosh^2 r\left[{\ell(\ell+1)a\over 2}H^{(+)\ell m}_{n}
   +\tanh r~ w^{(+)\ell m}\right],
\end{equation}
and $\hat K$ is the derivative operator defined by 
\begin{equation}
 \hat K = -a\partial_{T}a^3\partial_{T}a^{-2}.
\end{equation}
To keep the simplicity of notation, we often 
abbreviate the indices, $(\pm),\ell$ and $m$, unless there arises 
confusion.

Then we can see easily that $w$ and $\Pi$ can be expanded 
in terms of the eigen function of the operator $\hat K$. 
The normalized eigen functions should satisfy 
\begin{equation}
 \hat K {\cal U}_p (T)=(p^2+1) {\cal U}_p (T),
\label{D11}
\end{equation} 
and 
\begin{equation} 
 \int_{T_L}^{T_R}{dT \over a^3}{\cal U}_p\overline{{\cal U}_{p'}}
  =\left\{
  \begin{array}{cc}
  \delta(p-p'),& (p^2\ge 0), \\ 
  \displaystyle \sum_n \delta_{p,p_n}\delta_{p',p_n},& (p^2< 0).  
  \end{array}\right.
\label{D12}
\end{equation}
where $p_n^2+1$ is a discrete eigenvalue of the operator 
$\hat K$. 
We expand the variables $w$ and $\Pi$ as
\begin{equation}
 w^{(+)\ell m}=-\int dp~w_{(+)p\ell m} {\cal U}_p,\quad 
 \Pi^{(+)\ell m}
   =-\int dp~\Pi_{(+)p\ell m} {\cal U}_p.
\end{equation}

As we adopt the synchronous gauge condition, we can 
write down the mode functions by using the tensor harmonics as 
\begin{equation}
 h_{ij}=\sum {\cal N}_{(+)pl\, m}{\cal U}_{p}(T)
 {\cal G}_{ij}^{(+)pl\, m}, 
\label{D14}
\end{equation}
where ${\cal N}_{(+)pl\, m}$ is a normalization 
constant to be determined later. 
Then comparison of the traceless part of $h^{(+)p\ell m}_{AB}$ with
the definition of $w$ in Eq.~(\ref{vardef}) readily gives
 the solution for $w_{(+)p\ell m}(r)$,
\begin{equation}
w_{(+)p\ell m}
=
    -{{\cal N}_{(+)p\ell m}\over 2}{\cal P}_{p\ell}\,,
\label{A38}
\end{equation}
which, of course, satisfies the equation of motion which follows 
from the reduced action (\ref{redact}). 
Then repeating the same discussion succeeding to Eq.~(A37) of
Paper I,
the normalization is found to be fixed as\footnote{
The normalization of $\cal P$ is different from that in
Paper I by the factor of $\sqrt{2}/\Gamma(ip+l+1)$.}
\begin{equation}
 {\cal N}_{(+)pl\, m}=\sqrt{(l-1)l(l+1)(l+2)\Gamma(ip+l+1)
\Gamma(-ip+l+1)\over 4 p^2(p^2+1)}. 
\label{nconst}
\end{equation}

For $p^2>0$, taking account of the fact that the normalization of 
${\cal U}_p$ defined by Eq.~(\ref{D12}) gives rise to an additional
factor $2/|\Gamma(ip)|^2$ when their Klein-Gordon norms are evaluated on
hypersurfaces in $R$ and $L$ (see Appendix A of Ref.~\citen{STY95}),
we can see that the same normalization 
is deduced from the action (\ref{2action}).

As for $p^2<0$, the normalization constant (\ref{nconst}) is finite
except for the case $p^2=-1$. 
Hence the supercurvature modes would exist if there were modes
of ${\cal U}_p$ that would satisfy the normalization condition
(\ref{D12}). But we know that such modes do not exist from the
discussion of Appendix C. 

In the case $p^2=-1$, the normalization constant ${\cal N}_{(+)pl\,m}$
diverges.
On the other hand, since ${\cal U}_p\propto a^2$ for $p^2=-1$, the
integral,
\begin{equation}
 I_p^2=\int_{T_L}^{T_R} {dT \over a^3} \vert{\cal U}_p\vert^2 ,
\label{Ip2}
\end{equation}
is finite. Hence the overall normalization factor, which is given by
${\cal N}_{(+)pl\, m}/I_p$, diverges. This implies the $p^2=-1$ modes 
are `zero modes' for which the potential in the configuration space
along the direction of the modes is flat.
As shown in Appendix B, this corresponds to the fact that
the $p^2=-1$ modes are gauge illusion
and there is no dynamical degree of freedom there. So we do not have 
worry about this case. 

Another case of the divergent ${\cal N}_{(+)pl\,m}$
occurs at the boundary between the subcurvature and supercurvature
modes, $p^2=0$. In this case, however, an analysis of the asymptotic
behavior of ${\cal U}_p$ as $T\to T_J$ ($J=L,R$) shows
the integral (\ref{Ip2}) diverges as well.
Therefore the overall normalization factor 
${\cal N}_{(+)pl\, m}/I_p$ becomes indefinite as $\infty/\infty$
and we cannot conclude there is no physical discrete mode there.
This case is special in the sense that the 
even parity $p^2=0$ tensor 
harmonics can be constructed from the $p^2=-4$ scalar
harmonics\cite{HAMA,Garriga,YST96}. 
Thus these modes are degenerate with scalar type perturbations in 
the language of cosmological perturbation theory and a complete
treatment can be done only if the perturbation of the scalar field is
taken into account at the same time. Such an analysis has been given
in Appendix D and it has been shown that there exists no physical
discrete mode at $p^2=0$.

\subsection{odd parity}

We expand the metric perturbation in terms of the spherical harmonics as
\begin{equation}
 h^{(-)}_{nA}=\sum H^{(-)\ell m}_{n}{\cal Y}_{A},\quad  
 h^{(-)}_{\xi A}=\sum H^{(-)\ell m}_{\xi}{\cal Y}_{A},\quad
 h^{(-)}_{AB}=\sum w^{(-)\ell m} {\cal Y}_{AB}\,.
\end{equation}
As in the even parity case, the same argument below Eq.~(A49) of
Paper I 
holds with the replacements listed in Eq.~(\ref{replrule}). 

{}Finally we obtain the reduced action:
\begin{eqnarray}
 \int dr \int &&dT~{\cal L}^{(-)}_{(red)}
  =\sum_{\ell,m}{2\ell(\ell+1)\over (\ell-1)(\ell+2)}
    \int dr \int {dT\over a^3} 
   \Biggl[
    \Pi \hat K \left(\partial_r Q\right)
\cr && -{1\over 2}\left({1\over\cosh^2 r}\overline{\Pi}
   \hat K \Pi+
    \overline{Q} \hat K
    \left\{\ell(\ell+1)+\hat K\cosh^2 r\right\} Q\right)\Biggr], 
\end{eqnarray}
where
\begin{eqnarray}
  Q^{(-)\ell m}&:=&a H^{(-)\ell m}_n
\cr
 \Pi^{(-)\ell m}&:=&-{\ell(\ell+1)-2\over 2}w^{(-)\ell m} 
     -2a \sinh r\cosh r H^{(-)\ell m}_{n}.
\end{eqnarray}

As before, the variables $Q$ and $\Pi$ are expanded as
\begin{equation}
 Q^{(-)\ell m}=-a^2\int dp~Q_{(-)p\ell m} {\cal U}_p,\quad 
   \Pi^{(-)\ell m}
   =-a^2\int dp~\Pi_{(-)p\ell m} {\cal U}_p\,.
\end{equation}
We write down the mode functions by using the tensor harmonics as 
\begin{equation}
 h_{ij}=a^2\sum {\cal N}_{(-)pl\, m}{\cal U}_{p}(T)
 {\cal G}_{ij}^{(-)pl\, m}. 
\end{equation}
Then comparison of $h^{(-)p\ell m}_{r A}$ with
the definition of $Q$ gives
\begin{equation}
Q_{(-)p\ell m}
=
    -{{\cal N}_{(-)p\ell\,m}}{\cal P}_{p\ell}\,,
\end{equation}
Then as in the case of even parity modes 
the normalization is fixed as 
\begin{equation}
 {\cal N}_{(-)pl\, m}=\sqrt{(l-1)(l+2)\Gamma(ip+l+1)\Gamma(-ip+l+1)
\over 4 l(l+1)(p^2+1)}. 
\end{equation}

Different from even parity modes, the only exceptional case is 
the $p^2=-1$ modes. 
But again they are unphysical as shown in Appendix B.
Thus there exists no physical discrete modes.

\end{document}